\documentclass[12pt]{article}
\renewcommand{\baselinestretch}{1.3}
\usepackage{graphicx}
\usepackage{fullpage,amsmath,amssymb,latexsym}
\usepackage{multirow}
\usepackage{rotating} 
\usepackage{amsmath}
\usepackage{url}
\usepackage{graphicx}
\usepackage{caption}
\usepackage{subcaption}

\begin{document}

\title{A possible correlation between planetary radius and orbital period for small planets}
\author{Ravit Helled, Michael Lozovsky \& Shay Zucker\\
 Department of Geosciences,  
Tel-Aviv University, Tel-Aviv, Israel.}
\date{}
\maketitle

\begin{abstract}
We suggest the existence of a correlation between the
planetary radius and orbital period for planets with radii smaller
than 4 $R_{\oplus}$. Using the {\it Kepler} data, we find a correlation coefficient of 0.5120, and suggest that the correlation is not caused solely by survey incompleteness.  
While the correlation coefficient could change depending on the statistical analysis, the statistical significance of the correlation is robust. 
Further analysis shows that the correlation originates from two contributing factors. One seems to be a power-law dependence between the two quantities
for intermediate periods (3--100 days), and the other is a dearth of planets with radii larger than 2 $R_{\oplus}$ in short periods. 
This correlation may provide important constraints for
small-planet formation theories and for understanding the dynamical
evolution of planetary systems. 
\end{abstract}
\maketitle

\section{Introduction}
The {\it Kepler} mission (NASA) has detected thousands of planet candidates and has provided the opportunity to study planets statistically as a class of astrophysical objects.  
{\it Kepler} has indeed provided clues about the occurrence rate and physical properties of small planets with orbital periods up to about 500 days (e.g., Marcy et al. 2014a,b).  
\par

Several studies have been dedicated to the investigation of the physical properties of exoplanets (e.g., Lissauer et al. 2011; Lopez et al. 2012; Marcy et al. 2013; Rogers 2015) and of  the relations among their physical and orbital properties (e.g., Fischer \& Valenti 2005; Mazeh \& Zucker 2003 ; Guillot et al. 2006; Burrows et al. 2007; Mayor et al. 2011; Miller \& Fortney 2011; Buchhave et al. 2012; Howard et al. 2012). 
In this short Letter we investigate the
correlation between 
 planetary radius and orbital period for the {\it Kepler} candidates.  
We
focus on planets with radii smaller than 4 $R_{\oplus}$, i.e., Neptune-size planets and smaller, planets that are now known to be most common around other stars. 

\section{Statistical Analysis} 
We use all the {\it Kepler} candidates with radii up to  4 $R_{\oplus}$ and with orbital periods between 0.5 and 500 days. The choice of 4 $R_{\oplus}$ is based on the standard division of small planets and giant planets (e.g., Weiss \& Marcy 2014; Marcy et al. 2014b). We choose this orbital period range in order to include most of the data. 
Figure 1 shows the planetary radius ($R_p$) as a function of orbital
period ($P$) for our sample. The plot seems to suggest that there is a correlation between
the two parameters, and indeed, the correlation coefficient between $\log R_p$ and $\log P$ turns out to
be 0.5120. However, it is naturally harder to detect planets with
small radii at longer orbital periods,
which is a selection effect that already introduces some
correlation. Therefore, we must make sure that the correlation is not
simply caused by this selection effect. In order to quantify this influence, we use the completeness values for {\it Kepler} planet
detection, as derived by Silburt et al.~(2015, see their Figure 3). The 
 curve corresponding to a completeness value of
 $80\%$ is also shown in the Figure. Applying this completeness criterion leaves us
 with a sample of 2,955 planets (the blue dots above the sloped line in the Figure, henceforth the 'full sample').

In order to quantify the statistical significance of the correlation
we perform a bootstrap test (e.g., Efron \& Tibshirani 1993) in which we randomly draw a new sample
of $(P,R_p)$ pairs from the two separate samples of $R_p$ and $P$.
We exclude pairs which do not meet the completeness criterion
and calculate the correlation coefficient once this random sample
reaches the size of the original (full) sample.
This procedure leaves the marginal distributions of $R_p$ and $P$
essentially unchanged, it correctly  accounts for the effect of the completeness criterion, and it ruins any residual
correlation that is not caused by the incompleteness.
We repeat this resampling procedure $10^6$ times and in this
way obtain the null distribution of the correlation coefficient, against which 
we can test the hypothesis of the existence of correlation. 
\par

Out of $10^6$ random resamplings, none of the cases yields a correlation coefficient higher than that obtained for the true data. 
Figure 2 presents the distribution of the resampled correlation coefficients. This distribution is centred around a positive value of $\sim0.18$, and not zero, which reflects the effect of the completeness criterion. 
The arrow in the plot represents the correlation coefficient obtained for the true data (0.5120).  Since none of the random resamplings produced a value higher than the actual correlation, we can conclude that the correlation is statistically significance with a $P$ value smaller than $10^{-6}$.  

As can be seen from Figure 1, the {\it Kepler} candidates are not distributed uniformly in the radius-period diagram, and most of them are concentrated in the region between $\sim3 $ and 100 days. To understand which region dominates the correlation we  further divide the sample into three regions that exhibit somewhat different behaviors: (I) candidates with short periods of 0.5-3 days (II) candidates with intermediate periods, between 3 and 100 days, and (III) long-period candidates with periods between 100 and 500 days. The different regions are delineated in Figure 1. We repeat the analysis for each region separately including the bootstrap procedure.
Table~1 summarizes the derived correlation coefficients and corresponding $P$ values. 

Region (II) seems to dominate the correlation, and indeed Figure 1 qualitatively suggests a slope of ~0.5--0.6 in the log-log plane. However, a close examination of Figure 1 suggests an additional contributing factor, which is the lack of large planets in short orbital periods in region (I). 
We therefore divide region (I) into two sub-regions: (Ia) planets smaller than 2 $R_\oplus$; (Ib) planets larger than 2 $R_\oplus$.
There are 348 planets in region (Ia) and 62 in region (Ib). The areas of the two regions in the $\log P$-$\log R_p$ plane are 0.3847 (Ia) and 0.3010 (Ib).
Assuming, as our null hypothesis, a uniform distribution of the planets in this plane, the number of planets in (Ia) should follow a binomial distribution with $N=410$ and $p=0.5610$.
Under this null hypothesis the probability to obtain a number of 348 (or more) in region (Ia), is around $2\times10^{-36}$ -- undisputedly significant. 

\section{Discussion}
Our analysis and its results suggest that there is a correlation
between the planetary radii and the orbital periods for 
 planets that are smaller than 4 $R_{\oplus}$. If indeed true, this correlation implies
that larger planets are more likely to exist at larger
radial distances. 

The correlation follows from two contributing factors. The first is a correlation
between $R_p$ and $P$ for intermediate-size planets. Figure 1 suggests
a power-law relation of $R_p \sim P^{0.5-0.6}$. The second 
contributing factor is a depletion of large planets with short periods. 
This depletion has been noticed already by different authors (e.g., Ikoma \& Hori 2012; Owen \& Wu 2013; Lopez et al. 2013; Ciardi et al. 2013; Wu \& Lithwick 2013) and is most likely a signature of photo-evaporation of close-in
planets with atmospheres. 
This process naturally results in naked cores and small radii for planets that orbit very close to their stars. 
\par

It should be noted, that the analysis we present here does not consider the uncertainties of the radii and periods. While the errors of the periods are usually negligible, one might still wonder whether the relatively large errors of the radii could affect our findings. This could have been the case, if there were any correlation between those errors and the periods. We have calculated this correlation (not shown) and reassuringly found out it was negligible (less than 0.001). As a result, we can conclude that the correlation is robust and should not be affected by the measurement uncertainties. Future studies that account for a different sample of planets and perform a different statistical analysis could lead to a different value of the correlation coefficient but the correlation is expected to persist. 

Theoretical investigations are required to fully understand
the correlation and how it can constrain planet formation theories and
the evolution history of planetary systems. 
Future data from space
missions dedicated to detect transiting exoplanets such as {\it TESS,
CHEOPS} and {\it PLATO 2.0} and also ground-based observations,  
can be used to confirm and sharpen our results.

\subsection*{Acknowledgments} 
We thank the anonymous referee for valuable comments that helped to significantly improve this paper. R.~H.~acknowledges support from the Israel Space Agency under grant 3-11485. 

\renewcommand{\baselinestretch}{.9}

\newpage
 \begin{figure}
\vspace{-2. cm}
\centerline{\includegraphics[angle=0, width=12.cm]{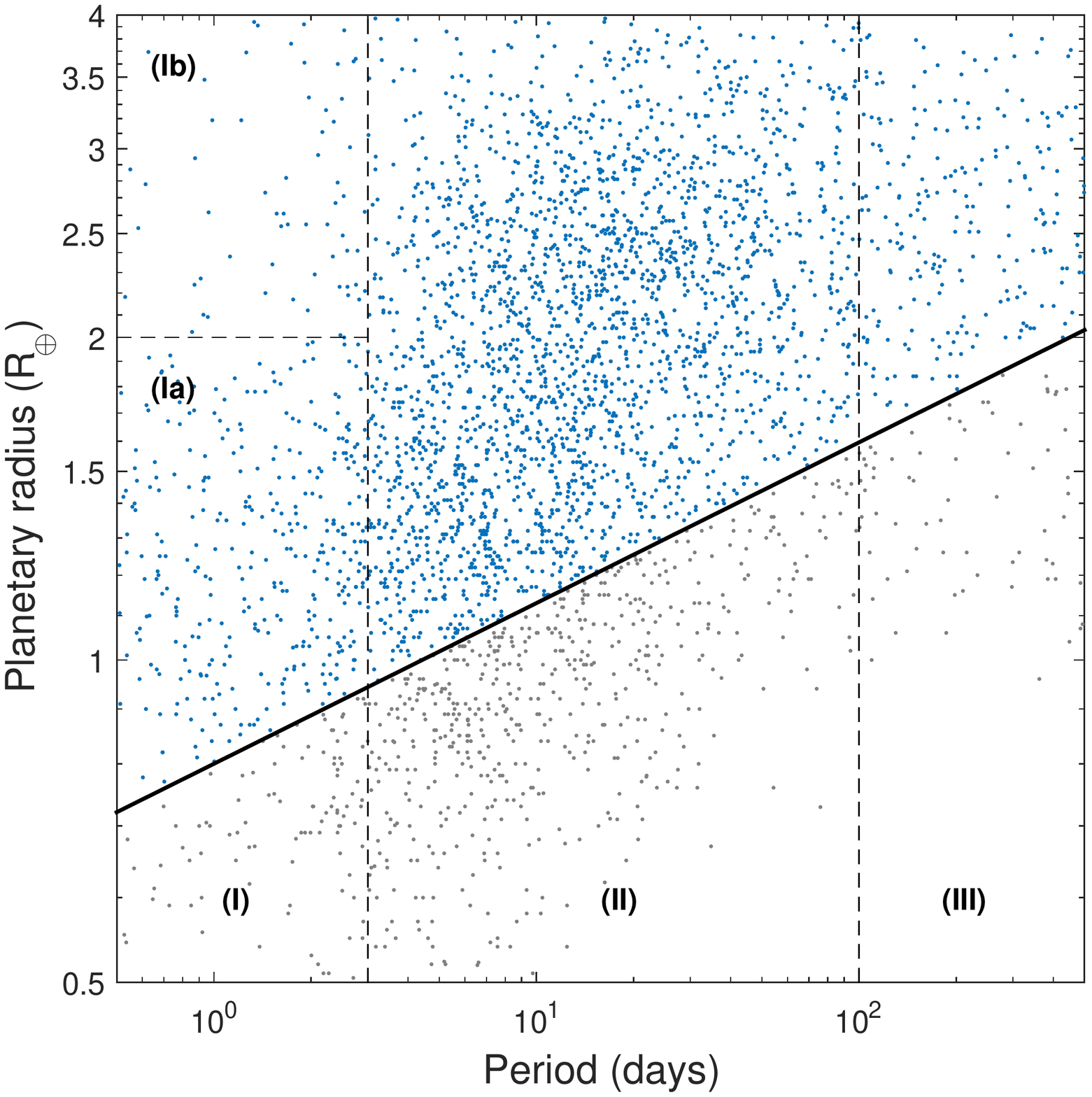}}
\vspace{-2. cm}
\caption{Planetary radius as a function of orbital period. The solid curve of $R_p = 0.8 \times P^{0.15}$ represents the 80\% {\it Kepler} data completeness curve. Also shown are the different region boundaries as discussed in the text. }
\end{figure}

\newpage
\begin{table}
\centering
{\renewcommand{\arraystretch}{1}
\begin{tabular}{|c|c|c|}
\hline
 & Correlation coefficient & $P$ value\\
\hline
Full sample & 0.5120 & $<10^{-6}$\\
Region (I) & 0.1583 & 0.0019 \\
Region (II)  & 0.3859 & $<10^{-6}$\\
Region (III)  & 0.0933 & 0.3699\\
\hline
\end{tabular}
}
\caption{{\small The calculated correlation coefficients and $P$ values for the different regions (see text for details).}}
\end{table}

\begin{figure}
\vspace{-2. cm}
\centerline{\includegraphics[angle=0, width=12.cm]{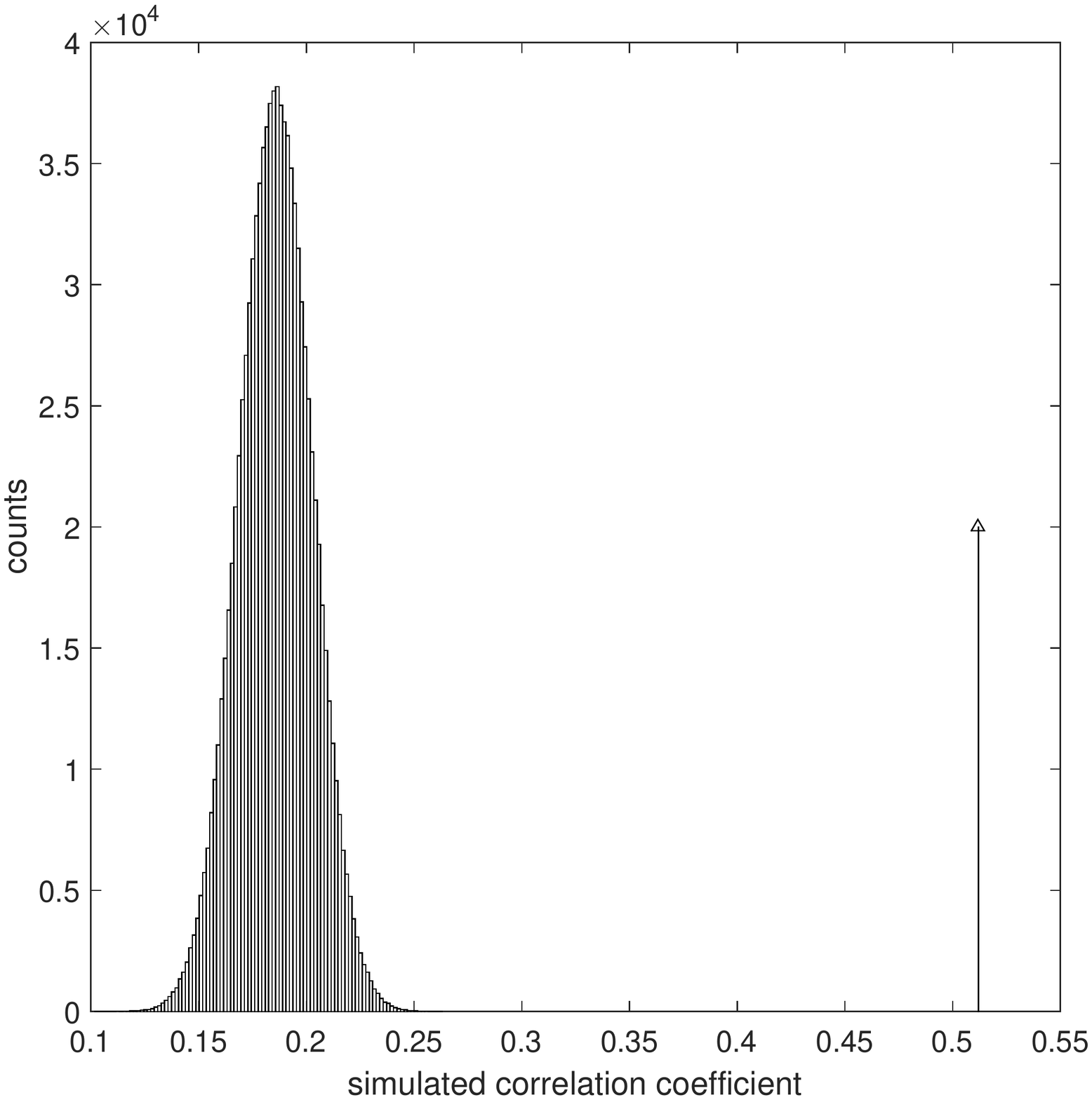}}
\vspace{-1.8 cm}
\caption{The distribution of $10^6$ resampled correlation coefficients (see text for details). The arrow shows the correlation coefficient computed for the true data. }
\end{figure}

\end{document}